\documentclass[epj]{svjour}
\usepackage{epsfig,epsf}

\usepackage{epsfig}
\usepackage{amsfonts}
\usepackage{amssymb}
\usepackage{amsmath}

\begin{document}
\title{Directed motion of domain walls in biaxial ferromagnets under
the influence of periodic external magnetic fields}
\titlerunning{Directed motion of domain walls in biaxial ferromagnets}
\author{Y. Zolotaryuk\inst{1}
 \and M.~M. Osmanov\inst{2}}
\institute{Bogolyubov Institute for Theoretical Physics,
National Academy of Sciences of Ukraine,
 03680 Kyiv, Ukraine, 
\email{yzolo@bitp.kiev.ua} 
\and 
National Technical University ``KPI'', 
Peremohy Av. 37, 03056 Kyiv, Ukraine,
\email{max.osmanov@gmail.com}
 }

\date{Received: date / Revised version: date}

\abstract{
Directed motion of domain walls (DWs) in a classical biaxial ferromagnet 
placed under the influence of periodic unbiased
external magnetic fields is investigated. Using the 
symmetry approach developed in this article the necessary 
conditions for the directed DW motion are found. 
This motion turns out to be possible if 
the magnetic field is applied along the easiest axis.  
The symmetry approach prohibits the directed DW
motion if the magnetic field is applied along any of the hard axes. 
With the help of the soliton perturbation theory and numerical
simulations, the average DW velocity as a function of different system
parameters such as damping constant, amplitude, and
frequency of the external field, is computed. 
}

\PACS{
{05.45.Yv}{Solitons} \and
{75.10.Hk}{Classical spin models} \and
{75.78.Fg}{Dynamics of domain structures}
}

\maketitle

\section{Introduction}
\label{sec1}

One-dimensional ferromagnetic models are currently of considerable
experimental and theoretical interest  \cite{kik90pr,ms91adp}. A large 
portion of this interest has been directed towards the domain wall (DW)
response to an ac magnetic field. One of the important problems here 
is the development of different ways for obtaining a net DW drift under 
the influence of unbiased perturbations.

The ratchet effect \cite{jap97rmp,r02pr,hmn05annpl,hm09rmp} has been shown to be 
an efficient tool to control the motion of particles and particle-like 
excitations. The mechanism of this effect is based on the breaking of 
all symmetries that connect two solutions with specular 
velocities \cite{fyz00prl}. For topological solitons this phenomenon 
has been investigated both 
theoretically \cite{m96prl,sq02pre,sz02pre,fzmf02prl,m-mqsm06c,zs06pre} and 
experimentally \cite{cc01prl,uckzs04prl,bgnskk05prl} in
continuous and discrete Klein-Gordon-type systems. It has been
shown \cite{sz02pre,fzmf02prl} that a {\it biharmonic} external 
field, consisting of a sinusoidal
signal and its {\it even} overtone can yield a directed soliton
motion. Similar biharmonic external field has been used in 
\cite{fo01pa} to control the dynamics of an individual spin. 

It should be mentioned that spatial asymmetry can be used for 
controlling the domain wall motion.
This can be achieved by creating a sawtooth-like asymmetric pattern on the
magnetic film \cite{srn05njp,srn06prb}. Experimental observation of this
phenomenon was reported in \cite{hokonms05jap}. Asymmetric pinning potential
that consists of triangular holes has been proposed and experimentally
implemented in \cite{ap-jr-rvmkaapm09jpd}. Observations have shown that 
this asymmetry favours certain direction of the domain wall propagation.
 
On the other hand, since the work of Schlomann \cite{sm74ieeetm,s75ieeetm} 
the problem of a DW drift under the influence of an oscillating magnetic field,
polarised either in the plane containing the easy 
axis \cite{bgd90jetp,clp91epl,k07pmm} or in the plane perpendicular to 
it \cite{sm74ieeetm,s75ieeetm,bgd90jetp}, has been
studied in the literature. Despite a certain number of papers devoted
to this problem, 
an interesting and important question arises: what are the
necessary conditions which one has to impose on the unbiased external
periodic magnetic field, such that a unidirectional DW motion will arise as
a result? Also, we would like to point out the need of a unifying approach
that would join together different ways of driving a unidirectional 
DW motion. In this paper, we show that the symmetry 
approach \cite{sz02pre,fzmf02prl} should be a perfect tool for this task.

Thus, the aim of this work is to investigate in detail the 
possibility of the unidirectional motion of magnetic topological 
solitons (domain walls). In particular, we formulate
the necessary conditions which have to be imposed on the external unbiased 
magnetic field in such a way that this motion will take place.

The paper is organised as follows. In the next section, we describe the
equations of motion for the biaxial ferromagnet. In Section 3,
the symmetries of the Landau-Lifshitz (LL) equation are discussed.
The average domain wall velocity is computed analytically in Section 4.
The numerical solution of the LL equation is given in Section 5. Conclusions
and a final discussion are presented in the last section.

\section{Model and equations of motion}
\label{sec2}

The dynamics of the one-dimensional chain of classical spins in the 
continuum limit is described by the well-known Landau-Lifshitz (LL)
equation 
\begin{equation}\label{1}
\partial_t {\bf S}=-[{\bf S}\times (\partial_x^2{\bf S} + {\hat J}{\bf S})]+\epsilon
{\bf f}({\bf S},t),
\end{equation}
where ${\bf S}(x,t)=(S_x,S_y,S_z)^T$ is a three-component
dimensionless magnetisation vector. Without
loss of generality, we assume the following normalisation condition:
$S_x^2+S_y^2+S_z^2=1$. The matrix
${\hat J}=\mbox{diag}(J_x,J_y,J_z)$ contains the information about the anisotropy
constants ($\beta_1\equiv J_x-J_y$, $\beta_3\equiv J_z-J_y$), so that 
the total energy of the magnet is given by 
\begin{equation}\label{energy} 
E=\frac{1}{2}\int_{-\infty}^{+\infty} [(\partial_x {\bf S})^2
-{\bf S}{\hat J}{\bf S}]~dx=
\int_{-\infty}^{+\infty} {\cal E}[{\bf S}(x,t)] dx.
\end{equation}
Note that if $\beta_1 <0$, $\beta_3>|\beta_1|$, the OZ axis is the
easiest axis, so that we have an easy axis ferromagnet with
XY being the anisotropic hard plane. 
Here ${\cal E}({\bf S})$ is the energy density function. The 
perturbative term $\epsilon {\bf f}$ contains 
the external magnetic field and the phenomenological Landau-Gilbert 
damping
\begin{equation}\label{2}
\epsilon {\bf f}=-[{\bf S} \times {\bf H}(t)]+\lambda [{\bf S}\times {\bf S}_t].
\end{equation}
Here  the periodic external magnetic field ${\bf H}(t)={\bf H}(t+T)$ has
zero mean value $\langle {\bf H}(t) \rangle_t=0$ and $\lambda$ is a 
damping constant. For the most of magnetic materials 
$\lambda \sim 0.008 \div 0.01$ \cite{aharoni}.

In this paper, we are interested in computing the average DW velocity
as a function of system parameters. But before embarking on this task,
it is necessary to investigate the symmetry properties of the LL
equation.

\section{Symmetries of the Landau-Lifshitz equation}
\label{sec3}

According to the previous work \cite{fzmf02prl,sz02pre}, we state that
the necessary condition for the occurrence of the directed DW motion
is the breaking of all the symmetries that relate two solitons with the same 

topological charge and with specular velocities:
\begin{equation}
\widehat{\cal S}\; {\bf S}(x,t;v)= {\bf S}(x,t;-v).
\end{equation}

The unperturbed ($\epsilon =0$) DW solution of the LL equation is
well known \cite{s79lomi}. It is a topological 
soliton ${\bf S}^{(0)}(x,t;\phi)$ $=(\cos {\phi}~ \mbox{sech}z,
~\sin {\phi}~ \mbox{sech}z,~Q \tanh z)$, where 
$z= \xi (x+ x_0) - Q \zeta t$=$\xi(x-x_0-vt)$, 
$\xi= \sqrt{\beta_3-\beta_1 \cos^2{\phi}}$,
$\zeta= \beta_1 \sin \phi~ \cos \phi$. Here $Q$ is the topological
charge of the DW soliton with $Q=1$ corresponding to the kink (soliton)
and $Q=-1$ to the antikink (antisoliton) solution. The azimuthal angle
 $\phi$ describes the direction of the projection of the magnetisation vector
${\bf S}$ on the XY plane, and thus the handedness
or polarity of the DW changes depending on the 
interval to which the value of $\phi$ belongs: $0<\phi<\pi$ or
$-\pi<\phi<0$. The DW velocity is defined by the
value of $\phi$, moreover, $v(\phi)=v(\phi+\pi)$ and $v(-\phi)=-v(\phi)$.

Taking into account the properties of the unperturbed soliton and 
assuming that
the perturbation is weak enough not to distort the soliton shape, one
can define the soliton velocity and the center of mass as
\begin{equation}\label{Xc}
v=\frac{dX_c}{dt} = \frac{1}{E}\int_{-\infty}^{+\infty} x 
\frac{\partial}{\partial t}
{\cal E}[{\bf S}(x,t)] dx.
\end{equation}
It is easy to see that there exist only two types of symmetries that 
can relate two arbitrary solutions with opposite velocities and the same 
topological charge. These operations must include either a space 
reflection and a time shift or, vice versa, a time reflection and a 
space shift:
\begin{eqnarray}\nonumber
\widehat{\cal S}_{1\alpha}:&& x \to -x+x', t \to t+ \tau_\alpha, 
S_{\alpha} \to -S_\alpha,\\
&& S_{z} \to -S_z;~~ \alpha=x,y, \label{S1}\\ 
\nonumber
\widehat{\cal S}_{2\alpha}:&& x \to x+x', t \to -t+ t', 
S_{\alpha} \to -S_\alpha; \\
&& \alpha=x,y, \label{S2}
\end{eqnarray}
where $x'$ and $t'$ are arbitrary constants and $\tau_\alpha=nT/2$, $n=0,1,2$. 
For the sake of clarity, let us briefly discuss the symmetries 
${\widehat {\cal S}_{1x}}$
and ${\widehat {\cal S}_{1y}}$. The symmetry ${\widehat {\cal S}_{1x}}$ acts
on the unperturbed solution by turning a DW ${\bf S}^{(0)}(x,t;\phi)$ into
a solution ${\bf S}^{(0)}(x,t;\pi-\phi)$, while the symmetry 
${\widehat {\cal S}_{1y}}$ turns a solution with $\phi$ into a solution
with $-\phi$. In these cases, $v(\pi-\phi)=-v(\phi)$ and
 $v(-\phi)=-v(\phi)$, respectively. Therefore, the abovementioned 
 symmetries connect two DW solutions with opposite velocities, while the 
other DW properties, in particular, the topological charge $Q$ or the 
width $\xi^{-1}$, remain unchanged. Note that although 
the symmetry ${\widehat {\cal S}}_{1y}$
changes the polarity of the DW, we still consider these two solutions
as the same because the polarity is a local characteristic of a
DW, in contrast to the topological charge.

Application of the perturbation ({\ref{2}}) can or cannot destroy the above 
symmetries. Note that when the magnetic field is applied, the 
energy density (\ref{energy}) must be complemented with the term 
$-({\bf S},{\bf H}(t))$.
The symmetries $\widehat{\cal S}_{1\alpha}$ are present if there exists such
$\tau$ that the following equalities are satisfied for the magnetic 
field ${\bf H}(t)$:
\begin{eqnarray}\nonumber
H_\alpha(t+\tau_\alpha)=-H_\alpha(t),~H_\beta(t+\tau_\alpha)=H_\beta(t),~\\
 H_z(t+\tau_\alpha)=-H_z(t); ~~\alpha=x,y,~~ \beta\neq \alpha~.\label{CS1}
\end{eqnarray}
One should stress that for each symmetry $\widehat{\cal S}_{1\alpha}$ there
is a corresponding value of $\tau_\alpha$.
 
The symmetries $\widehat{\cal S}_{2\alpha}$ are always violated in the presence
of dissipation ($\lambda \neq 0$), however, if $\lambda=0$, they are present if
there exists such $t'$ that the following equalities take place:
\begin{eqnarray}\nonumber
H_\alpha(-t+t')=-H_\alpha(t),\; 
H_\beta(-t+t')=H_\beta(t),\\
H_z(-t+t')=H_z(t);~~\alpha=x,y,~\beta \neq \alpha  .\label{CS2}
\end{eqnarray}

Thus, in a general case $\lambda \neq 0$, in order to obtain the 
directed soliton motion one has to apply a magnetic field for 
which in both (for $\alpha=x$ and $\alpha=y$) the sets of the 
equalities (\ref{CS1}) at least one equation in each set does not hold. 
In the dissipationless case, one has to violate
both the sets (\ref{CS1}) and (\ref{CS2}). 

Consider now the oscillating magnetic field directed along one of 
the coordinate axes. If ${\bf H}(t)||$OZ, in order to break the symmetries, 
$\widehat{\cal S}_{1\alpha}$ one has to choose the respective component 
$H_z(t)$ in such a way that it satisfies the inequality $H_z(t+T/2) \neq -H_z(t)$.
In the cases ${\bf H}(t)||$OX and ${\bf H}(t)||$OY, one of the
 symmetries $\widehat{\cal S}_{1\alpha}$ will 
always be present. Indeed, if ${\bf H}(t)=(H_x(t),0,0)^T$, the 
symmetry $\widehat{\cal S}_{1x}$ is present if there exists such $\tau_x$ that
$H_x(t+\tau_x)=-H_x(t)$. Obviously, this equation holds if $\tau_x=T/2$.
The symmetry $\widehat{\cal S}_{1y}$ is present if there exists
such $\tau_y$ that the equality $H_x(t+\tau_y)=H_x(t)$ holds. 
Since the external magnetic field is periodic, this condition is
automatically fulfilled for $\tau_y=T$. Similarly, if we consider  
${\bf H}(t)=(0,H_y(t),0)^T$, the symmetry $\widehat{\cal S}_{1y}$ is
present if $H_y(t+T/2)=-H_y(t)$, but for the presence of 
$\widehat{\cal S}_{1x}$ it is sufficient to guarantee the periodicity of the
function $H_y(t)$. Thus, it is not possible to obtain a 
directed soliton motion by applying magnetic field only along OX or
OY axis, at least, for arbitrary small perturbation. On the 
other hand, it is possible to obtain a directed motion in the case 
${\bf H}(t)||$OZ, since one can violate the equality $H_z(t+T/2)=-H_z(t)$
by various choices of the magnetic field, for instance, by choosing it 
in the following biharmonic form :
\begin{equation}\label{mag-field}
{\bf H}(t)={\bf e}_z H(t) = {\bf e}_z [H_1 \cos (\omega t)+
H_2 \cos (m\omega t+\theta)],
\end{equation}
with $m=2,3,\ldots$ . 
If $H_2 \neq 0$ and $m$ is even, the abovementioned equality is 
always violated, while for odd $m$'s it is always satisfied.
In the dissipationless case $\lambda=0$, another set of symmetries,
namely ${\widehat {\cal S}_{2\alpha}}$, must be broken. For a given
choice of the magnetic field direction, this situation occurs if there 
does not exist such $t'$ that $H_z(-t+t')= H_z(t)$. For the
function (\ref{mag-field}) this means that the symmetries 
${\widehat {\cal S}_{2\alpha}}$ are 
violated if $m$ is even and $H_2\neq 0$, $\theta \neq 0,\pm \pi$.

\section{Computation of the average DW velocity using the perturbation 
theory}

The perturbation theory in the first order implies that the 
perturbation (\ref{2}) is too small to distort the soliton shape and it 
influences only the temporal evolution of the following soliton parameters: 
its center of mass $X(t)$ and the azimuthal angle $\phi(t)$. 
The equations that describe the parameter evolution are obtained in 
accordance with the papers \cite{p86jetp,k89pd}:
\begin{eqnarray}
\frac{d\phi}{dt}&=&-\frac{\epsilon}{2} \int_{-\infty}^{+\infty}
\frac{f_-(z)}{\cosh z} dz, \\
\nonumber
\frac{dX}{dt} &=& \frac{Q \zeta }{\xi}-\frac{\epsilon}{2\xi}
\int_{-\infty}^{+\infty}\left [ 
\frac{\zeta}{\xi^2} \frac{zf_-(z)}{\cosh z}  +Q f_z\right ] dz,\\
f_{-}(z)&=&f_x \sin \phi-f_y \cos \phi,
\end{eqnarray}
where $f_\alpha$ ($\alpha=x,y,z$) are the components of the
perturbation vector (\ref{2}), and the soliton parameters 
$\xi(t)= \sqrt{\beta_3-\beta_1 \cos^2{\phi(t)}}$ and 
$\zeta(t)= \beta_1 \sin \phi(t) \cos \phi(t)$ now depend on time.

Consider the case of the magnetic field (\ref{mag-field}) directed 
along the OZ axis. In this case, the equations for the time evolution of 
the soliton parameters  $X(t)$ and $\phi(t)$ have the following form:
\begin{eqnarray} 
\label{teq1} \frac{d\phi}{dt}&=&H(t)+\Gamma \sin 2\phi,
~~\Gamma=-\frac{\lambda \beta_1}{2},\\ 
\frac{d X}{dt}&=& -Q \frac{\beta_1 \sin 2\phi }{2\sqrt{\beta_3-
           \beta_1 \cos^2 \phi}}.
\label{teq2}
\end{eqnarray}

If one assumes that the soliton shape changes only insignificantly
due to the perturbation, the vibrations of the azimuthal angle $\phi(t)$ 
around its equilibrium position appear to be small. Next, we assume the 
amplitudes $H_{1,2}$ to be small parameters. Then the oscillating 
solution of the equation (\ref{teq1}) should be sought in the
form 
$\phi(t)=\pm \pi/2+H_1 \phi_1(t)+H_1^2 \phi_2(t)+H_1^3 \phi_3(t)+{\cal O}(H_1^4)$.
The initial value $\pm \pi/2$ comes from the fact that the Bloch
wall configuration is energetically most favourable. 
After representing the external magnetic field in the form 
$H(t) \equiv H_1 h(t)= H_1 [\cos {\omega t}+ H_2/H_1 \cos {(2\omega t+\theta)}]
$ and substituting the expansion for $\phi(t)$ into equation (\ref{teq1}), 
one finds an approximate expression for $\phi$ which contains only
the $\phi_1(t)$ and $\phi_3(t)$ terms. Then it remains only to substitute
the expression for $\phi(t)$ into equation (\ref{teq2}) and to average it 
over one oscillation period. In the expansion of the r.h.s of the
equation (\ref{teq2}) into the Taylor series with respect to the parameter $H_1$, 
we have limited ourselves with the term of the order ${\cal O}(H_1^3)$. As a
result, the following expression for the average DW velocity is obtained:
\begin{eqnarray}\nonumber
&&\langle v \rangle   \simeq 
 -QA(H_{1,2},\omega,\Gamma)  
\sin(\theta-\theta_0),\\
&& A(H_{1,2},\omega,\Gamma)=\frac{3\beta_1^2}{16\beta_3^{3/2}}
\nonumber
\frac{H_2H_1^2}{(4\Gamma^2+\omega^2)\sqrt{\Gamma^2+\omega^2}},\\
&&\theta_0=2 \arctan(2\Gamma/\omega)-\arctan(\Gamma/\omega).\label{v1} 
\end{eqnarray}
We would like to stress that this expression is the same  for   
the initial angles $\phi=\pi/2$ and $\phi=-\pi/2$.

In the case of odd $m$'s, we obtain $\langle v \rangle=0$. The 
expression (\ref{v1}) clearly confirms the validity of the
symmetry approach. The average soliton velocity becomes zero
if $H_2=0$ and this signals the restoration of the symmetries 
${\widehat {\cal S}_{1\alpha}}$. In the dissipationless limit 
($\lambda \to 0$), $\theta_0 \to 0$ and thus 
$\langle v \rangle \propto \sin \theta$. In this limit, the
symmetries ${\widehat {\cal S}_{2\alpha}}$ are restored if
$H(t)=H(-t)$. The average DW velocity becomes zero at the values 
$\theta=0,\pm \pi$, so that they are precisely those values at which 
the function $H(t)$ is symmetric.

\section{Numerical simulations}
\label{sec4}

In order to verify the symmetry approach developed in Section \ref{sec3},
the initial LL equation has been discretized in the spatial dimension 
with the step $h=0.05$ and the resulting system of coupled ordinary
differential equations has been integrated numerically using the fourth 
order Runge-Kutta method. It is convenient to solve numerically the
LL equation in the following form:
\begin{eqnarray} \nonumber
&& -\partial_t {\bf S}=\frac{1}{1+\lambda^2} \left [{\bf S} 
\times {\bf H}_* \right ] +
\frac{\lambda}{1+\lambda^2} \left [{\bf S} \times \left [{\bf S} \times 
{\bf H}_* \right] \right],\\
&&{\bf H}_*=\partial^2_x {\bf S} + {\hat J} {\bf S}+ {\bf H}(t),
\label{LL2}
\end{eqnarray}
which is equivalent to equation (\ref{1}) with the perturbation (\ref{2}). 
The validity of this method has been checked by monitoring the energy 
conservation in the purely Hamiltonian case $\lambda=0$ and
 ${\bf H}(t)=0$.

It should be emphasized that in order to compute the mean DW
velocity, one has to average over the set of initial conditions:
 phase $-\pi \le \phi\le \pi $, initial time $0<t_0<T$, and time $t$. 
In the numerical simulations, we consider only the dissipative 
case $\lambda>0$, therefore
we are interested in attractor(s) that correspond to moving DWs.
If the perturbation (\ref{2}) is small, then the phase space of the
LL equation have to consist of the basin(s) of attraction of 
{\it periodic} attractor(s) [limit cycle(s)] that are locked to the
frequency $\omega$ of the external magnetic field. Breaking the
respective symmetries should manifest itself in {\it desymmetrization}
of the basins of attraction that correspond to DWs moving with
opposite velocities. Below we demonstrate that actually in the case
of broken symmetries there exists only one attractor that corresponds
to the directed DW motion.  
In this case, it is sufficient to compute only the average velocity on
the attractor: $\langle v \rangle \to_{t \to \infty} [X(t+T)-X(t)]/T$,
where $X(t)$ is the DW center of mass. 

\begin{figure}[htb]
\centerline{\psfig{file=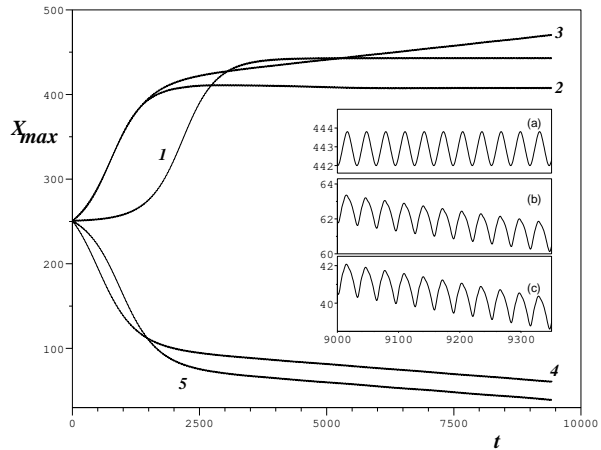,height=3.7in,angle=-90}}
\caption{Time evolution of the kink center $X_{max}$ under action of 
the external magnetic field (\ref{mag-field}) for $J_x=0.2$, $J_y=0.5$, 
$J_z=1$, $\lambda=0.01$, $\omega=0.2$. Curve 1 corresponds to the
case of single harmonic drive ($H_1=0.1$, $H_2=0$), curve 2
corresponds to $H_1=H_2=0.1$, $m=3$, $\theta=-2$. Curves 3-5 
correspond to $H_1=H_2=0.1$, $m=2$, $\theta=-1$ (curve 3),
$\theta=2.5$ (curves 4,5). The last two curves show evolution for
different initial times: $t_0=T/4$ (curve 4) and $t_0=0$ (curve 5).
Insets (a)-(c) give details of curves 1,4, and 5, respectively.} 
\label{fig1}
\end{figure}

First, we consider the case ${\bf H}(t)||$OZ and the expression for the
Z-component of the magnetic field given by equation (\ref{mag-field}). The 
time evolution of the DW center computed as   
$X_{max}=\mbox{max}_{x \in (-\infty,+\infty)}{\cal E}[{\bf S}(x,t)]$ 
(shown in Figure \ref{fig1}) clearly demonstrates the validity of 
the symmetry approach. 
%
Curve 1 in this figure corresponds to the case of
the single harmonic drive ($H_2=0$), where no directed DW motion is seen. 
Next, no directed DW motion is observed in the case of mixing two
odd harmonics ($m=3$) as shown by curve 2. Curves 3-5 illustrate the 
evolution of the DW center for the case of two mixed harmonics with 
$m=2$. It is easy to notice that on the time scale $t > \lambda^{-1}$, the 
system settles on a periodic attractor (the periodicity can be observed 
from the insets in Figure \ref{fig1}) that corresponds to the motion in the
direction, defined by the phase shift $\theta$. The simulations have
been performed for different initial values of $\phi$ and initial
times $t_0$, and in all the cases the system settles on the same 
attractor (compare, for example, curves 4 and 5).

\begin{figure}[htb]
\centerline{\psfig{file=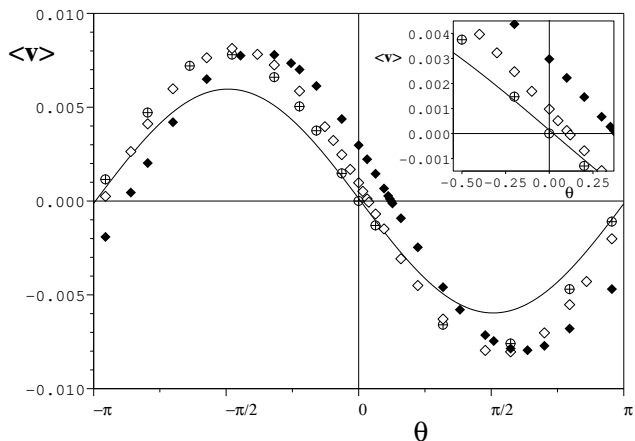,height=3.7in,angle=-90}}
\caption{Dependence $\langle v \rangle=\langle v \rangle(\theta)$
for $\lambda=0.01$ ($\oplus$), $\lambda=0.1$ ($\diamond$, scaled
by factor $1.3$), and $\lambda=0.3$ ($\blacklozenge$, scaled by 
factor $1.85$) computed numerically from equation (\ref{LL2}). The 
perturbation theory result (\ref{v1}) for $\lambda=0.01$ is shown 
by solid line. Other parameters are chosen the same as for curves 3-5 
in Figure \ref{fig1}. The inset shows details in the vicinity of $\theta=0$.
} 
\label{fig2}
\end{figure}

The dependence $\langle v \rangle (\theta)$ of the average DW velocity 
on the phase shift between the harmonics, $\theta$, is shown in 
Figure \ref{fig2}. 
%
It appears to have a sinusoidal shape, as predicted by
the perturbation theory result (\ref{v1}). 
Another demonstration of the validity of the symmetry approach is the
behavior of the points where $\langle v \rangle (\theta)=0$. 
In Figure \ref{fig2} along with the data for $\lambda=0.01$, the data
for $\lambda=0.1$ and $\lambda=0.3$ have been plotted as well. Although 
these values of damping do not correspond to realistic values for magnets,
these results are very instructive for the illustration of the 
restoration of the symmetries ${\widehat {\cal S}_{2\alpha}}$.
Indeed, when $\lambda \to 0$, the values of $\theta$, at which the 
DW velocity becomes zero, gradually shift to $0, \pm \pi$. As shown 
already in Section \ref{sec3}, both the symmetries 
${\widehat {\cal S}_{2\alpha}}$ are restored if $H_z(-t)=H_z(t)$.
This happens if $\theta=0,\pm \pi$ [see equation (\ref{mag-field})].
We would like also to point out good correspondence between the
perturbation theory results (shown by the solid line) given by 
equation (\ref{v1}) and the results of direct integration of the LL equation.

\begin{figure}[htb]
\centerline{\psfig{file=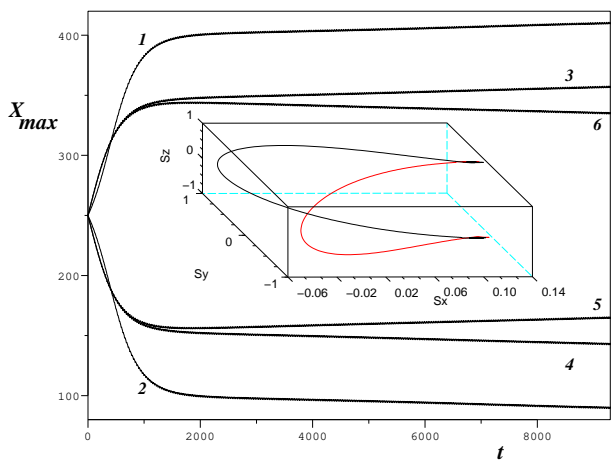,height=3.8in,angle=-90}}
\caption{ Time evolution of the kink center $X_{max}$ under 
action of the external magnetic field (\ref{mag-field}) 
for $H_1=H_2=0.05$, $\theta=1$, $m=2$ with initial azimuthal angle 
$\phi=\pi/10$ (curve 1), $\phi=-\pi/10$ (curve 2),
$\phi=\pi/4$ (curve 3), $\phi=-\pi/4$ (curve 4), 
$\phi=3\pi/4$ (curve 5), and $\phi=-3\pi/4$ (curve 6).
The rest of parameters is as in the case of Figure \ref{fig1}.
The inset shows distribution of the magnetisation 
vector components $(S_x, S_y, S_z)^T$ of DWs that correspond
to curves 3 (black) and 4 (red) of the main figure.} 
\label{fig3}
\end{figure}

Next, we consider the case of the magnetic field directed along the OX 
axis [${\bf H}(t)={\bf e}_xH(t)$], but which has the same functional 
dependence (\ref{mag-field}). According to the symmetry approach, this 
periodic drive does not yield the directed DW motion. The numerical 
simulations of this case are illustrated by Figure \ref{fig3}. The basins 
of attraction of two DW solutions that have opposite polarities and 
opposite velocities appear to be symmetric with respect to the value 
$\phi=0$. Indeed, if the initial value
of the azimuthal angle is positive (see curves 1,3, and 5), the 
dynamics of the system settles on the attractor with the positive DW
velocity ($v=0.00031$), while for negative values of $\phi$ (see 
curves 2,4 and 6) it tends to the solution with the 
DW velocity of opposite sign ($v=-0.00031$).
%
In the case of ${\bf H}(t)||$OY, the same scenario is observed.

\section{Discussion and conclusions}

In this paper, the symmetry approach for the analysis of the 
{\it unidirectional} domain wall motion has been developed. The main 
objective was to demonstrate that a proper choice of the oscillating 
unbiased magnetic field can yield a net directed motion of the domain 
wall. With the help of the 
symmetry approach, we have obtained the necessary conditions to be 
imposed on the magnetic field in order to obtain the unidirectional 
motion. When the magnetic field is applied along
a certain coordinate axis, this motion turns out to be possible only if this axis
coincides with the easy axis, say OZ. The symmetry arguments prohibit the directed DW
motion if the magnetic field is applied along any of the hard axes. 
The necessary condition for the directed DW motion in the 
presence of dissipation is given by $H_z(t)\neq -H_z(t+T/2)$. Next, we 
demonstrate the cubic 
dependence $\langle v \rangle \propto H_1^2H_2$ of the average DW
velocity on the magnetic field amplitude. These results appear to be
similar to those for topological solitons
in the ac driven sine-Gordon (SG) equation \cite{sz02pre}. Very recently,
 \cite{qca-n10pre} a rigorous proof of the universality of these
results for a wide range of nonlinear systems driven by the
biharmonic signals of the type (\ref{mag-field}) has been obtained.
 
However there are certain differences in the directed soliton  motion 
in the LL and SG cases. Due to the fact that the LL equation
is three-component, while SG is scalar, there are more ways to 
apply the external field to the system, namely along any of three 
coordinate axes. But only in the case of magnetic field applied
along the easy axis, the directed motion is possible. Here it should be 
stressed that we are interested in the average net motion which is 
independent from the 
the local properties of the DW such as handedness (polarity). The  
application 
of any ac signal along non-easy axes drives DWs with opposite polarities
into opposite directions. The basins of respective attractors of the LL
equation are symmetric with respect to the $\pi$-shift of the initial azimuthal
angle. Therefore, a weak noise which is inevitable in realistic systems will
lead to exploration of the whole phase space and eventually to zero net 
motion.

Another difference with respect to the SG case caused by the
multicomponentness of the LL equation is a wider range of possibilities
to drive a DW by external oscillating fields. Let us briefly 
discuss the case when two of the magnetic field components 
are nonzero. Consider first the case of magnetic field polarised
in the easy plane XY [${\bf H}(t)=(H_x(t),H_y(t),0)^T$]:
$H_x(t)=H_0^{(x)} \cos (m_x\omega t+\theta_x)$, 
$H_y(t)=H_0^{(y)} \cos (m_y \omega t+\theta_y)$. Here 
$m_x, m_y=1,2,3,\ldots$, and they do not have a common divisor.
Note that such a way to control the DW motion has been suggested
in \cite{bgd90jetp} for the particular 
case of $m_x=m_y=1$. The symmetry $\widehat{\cal S}_{1x}$ requires the 
simultaneous fulfilment of the
equalities $H_x(t+T/2)=-H_x(t)$ and $H_y(t+T/2)=H_y(t)$. Similarly, the 
$\widehat{\cal S}_{1y}$ symmetry is present if both the equalities 
$H_x(t+T/2)=H_x(t)$ and $H_y(t+T/2)=-H_y(t)$ hold. 
If both $m_x$ and $m_y$ are odd, we have 
$H_\alpha(t+T/2)=-H_\alpha(t) \neq H_\alpha(t)$. Therefore none
of the sets of equations (\ref{CS1}) can be satisfied and 
 thus both symmetries $\widehat{\cal S}_{1\alpha}$ are broken. 
If $m_x$ is even and $m_y$ is odd, we have $H_x(t+T/2)=H_x(t) \neq -H_x(t)$, 
 thus $\widehat{\cal S}_{1x}$ is broken. But, $H_y(t+T/2)=-H_y(t)$, 
therefore  $\widehat{\cal S}_{1y}$ is present. Similarly, if $m_y$ is 
even and $m_x$ is odd, the symmetry $\widehat{\cal S}_{1x}$ is satisfied 
and $\widehat{\cal S}_{1y}$ is broken.
Therefore if one of $m_{x,y}$ is even and another one is odd, we 
expect no directed motion, whereas this motion must occur if both $m_{x,y}$ 
are odd.

Another way to drive unidirectionally a domain wall is to apply 
the oscillating magnetic field polarised in the plane that contains the
easy axis \cite{k07pmm}:\\
$H_y(t)=H_0^{(y)} \cos (m_y\omega t+\theta_y)$,
$H_z(t)=H_0^{(z)} \cos (m_z \omega t+\theta_z)$; $m_y, m_z=1,2,3,\ldots$.
Alternatively, one can consider the field polarised in the XZ plane.
If both $m_y$ and $m_z$ are odd, it is impossible to break simultaneously
the set of equalities (\ref{CS1}) and (\ref{CS2}). In this case, the 
breaking of $\widehat{\cal S}_{1x}$ takes place because 
$H_z(t+T/2)=-H_z(t)$ but $H_y(t+T/2)=-H_y(t)\neq H_y(t)$. However, the symmetry
$\widehat{\cal S}_{1y}$ is still present. If $m_y$
 is even and $m_z$ is odd, we obtain 
$H_z(t+T/2)=-H_z(t)$ but $H_y(t+T/2)=H_y(t) \neq -H_y(t)$, therefore  
  $\widehat{\cal S}_{1x}$
 is present while  $\widehat{\cal S}_{1y}$  is broken.
If $m_y$ is odd and $m_z$ is even both the symmetries 
$\widehat{\cal S}_{1x,y}$ are broken because 
$H_z(t+T/2)=H_z(t) \neq -H_z(t)$. 
Therefore this is the only way
to obtain the directed DW motion with the help of magnetic
field polarised in the YZ plane.

Finally, we would like to outline the future directions of applications 
of the symmetry approach. It is of 
interest to consider topological magnetic excitations in two- and
three-dimensional systems, where alongside the directed translational
motion a unidirectional rotation can take place as well, as 
shown previously for particles \cite{dzfy08prl}. Another 
question is how to apply the symmetry approach to the problem of
DW directed motion in more complicated magnetic systems such as antiferromagnets,
ferrites, magnetoelastic systems, and others. In this
direction, some progress has already being accomplished in the papers
\cite{ggg-d94prb,gs95jmmm}.

\end{document}